\documentclass[12pt,a4paper]{article}
\usepackage{a4wide}
\usepackage{amssymb}
\begin{document}
{\renewcommand{\thefootnote}{\fnsymbol{footnote}}
\hfill  PITHA -- 99/27\\
\medskip
\hfill hep--th/9908170\\
\medskip
\begin{center}
{\LARGE Abelian $BF$-Theory and\\ Spherically Symmetric Electromagnetism}\\
\vspace{1.5em}
Martin Bojowald\footnote{e-mail address: 
  {\tt bojowald@physik.rwth-aachen.de}}\\
Institute for Theoretical Physics, RWTH Aachen\\
D--52056 Aachen, Germany\\
\vspace{1.5em}
\end{center}
}

\setcounter{footnote}{0}

\newtheorem{theo}{Theorem}
\newtheorem{lemma}{Lemma}
\newtheorem{defi}{Definition}

\newcommand{\proofend}{\raisebox{1.3mm}{\fbox{%
  \begin{minipage}[b][0cm][b]{0cm}\end{minipage}}}}
\newenvironment{proof}{\noindent{\it Proof:} }%
{\mbox{}\hfill \proofend\\\mbox{}}

\newcommand{\Ab}{\overline{{\cal A}}}
\newcommand{\Gb}{\overline{{\cal G}}}
\newcommand{\An}{{\cal A}_0}
\newcommand{\Ans}{{\cal A}_{0,\sigma}}
\newcommand{\Anb}{\overline{{\cal A}}_0}
\newcommand{\AnG}{{\cal A}_0/{\cal G}}
\newcommand{\AnGb}{\overline{{\cal A}_0/{\cal G}}}
\newcommand{\Lie}{{\cal L}}
\newcommand{\kt}{\vartheta}
\newcommand{\kp}{\varphi}

\newcommand{\md}{\mathchoice%
  {\mbox{\rm \,d}}%
  {\mbox{\rm \,d}}%
  {\mbox{\scriptsize\rm d}}%
  {\mbox{\tiny\rm d}}}
\newcommand{\td}[2][]{\mathchoice%
  {\frac{\mbox{\small\rm d}#1}{\mbox{\small\rm d}#2}}%
  {\frac{\mbox{\scriptsize\rm d}#1}{\mbox{\scriptsize\rm d}#2}}%
  {\frac{\mbox{\tiny\rm d}#1}{\mbox{\tiny\rm d}#2}}%
  {\frac{\mbox{\tiny\rm d}#1}{\mbox{\tiny\rm d}#2}}}
\newcommand{\pd}[2][]{\frac{\partial #1}{\partial #2}}
\newcommand{\fd}[2][]{\frac{\delta #1}{\delta #2}}
\newcommand{\Ad}{\mbox{\rm Ad}\,}
\newcommand{\ad}{\mbox{\rm ad}\,}
\newcommand{\tr}{\mbox{\rm tr}\,}
\newcommand{\sgn}{\mbox{sgn}}
\newcommand{\projlim}{\mbox{proj}\lim}
\newcommand{\Diff}{\mbox{Diff}}
\newcommand{\Hphys}{{\cal H}_{\mbox{\small phys}}}

\newcommand{\ScProd}[3][]{\langle#2,#3\rangle_{#1}}
\newcommand{\DirB}[3][\mkern-5mu]{\ensuremath{%
 \left\langle#2\left|#1\right|#3\right\rangle}}
\newcommand{\ket}[1]{\ensuremath{\left|#1\right\rangle}}

\newcommand*{\R}{{\mathbb R}}
\newcommand*{\Z}{{\mathbb Z}}
\newcommand*{\N}{{\mathbb N}}
\newcommand*{\C}{{\mathbb C}}

\begin{abstract}
  Three different methods to quantize the spherically symmetric sector
  of electromagnetism are presented: First, it is shown that this
  sector is equivalent to Abelian $BF$-theory in four spacetime
  dimensions with suitable boundary conditions. This theory, in turn,
  is quantized by both a reduced phase space quantization and a spin
  network quantization. Finally, the outcome is compared with the
  results obtained in the recently proposed general quantum symmetry
  reduction scheme. In the magnetically uncharged sector, where all
  three approaches apply, they all lead to the same quantum theory.
\end{abstract}

\section{Introduction}

Recently, H.\ Kastrup and the author proposed a general framework
for a quantum symmetry reduction procedure of diffeomorphism invariant
theories of connections \cite{SymmRed}. In the case of a reduction of
electromagnetism to its spherically symmetric sector an explicit
expression for the quantum symmetry reduction and for the observables
of the reduced theory was obtained. In the present paper we study
this sector in more detail by providing another approach to
symmetry reduction and quantization which, however, has the
drawbacks of being applicable for this special theory only and of
requiring a vanishing magnetic charge (or an explicit coupling to an
external one). Nevertheless, the methods involved are more standard
and this alternative quantization can serve as a simple test of the
general quantum symmetry reduction of Ref.~\cite{SymmRed}.

The new approach makes use of a novel identification of an Abelian
$BF$-theory \cite{Horo:BF} with the spherically symmetric sector of
electromagnetism. More precisely, it is proved that a partial gauge
fixing of an Abelian $BF$-theory with suitable boundary conditions is
equivalent to this symmetric sector upon a straightforward
identification of their variables. The two constraints of the
$BF$-theory provide the Gau\ss\ constraint of electromagnetism, and a
second constraint which on the one hand serves to perform after gauge
fixing a symmetry reduction of the theory and on the other hand
constrains the magnetic charge to vanish.

At first sight it may be surprising that we identify a sector of
electromagnetism with a topological field theory, but this causes no
problems because the kinematics of the spherically symmetric sector is
indeed diffeomorphism invariant, which leads to boundary observables
only (the electric charge and its conjugate momentum) after solving
the Gau\ss\ constraint. But the dynamics is not diffeomorphism invariant
because we need a background metric to construct the Hamiltonian.
Alternatively, we can couple the spherically symmetric sector to
gravitation, thereby rendering the metric dynamical and restoring
diffeomorphism invariance. This leads to an interpretation of the
spherically symmetric sector of electromagnetism as the
electromagnetic part of a Reissner-Nordstr\o m gravitational system
(which was our original motivation to study this sector). However, as the
gravitational degrees of freedom complicate the theory considerably,
we will not study their dynamics in this article.

In order to be able to ignore the gravitational degrees of freedom,
and at the same time maintain diffeomorphism invariance (which is
necessary to employ a spin network quantization) we regard the
electromagnetic sector as a spherically symmetric sector of
electromagnetism coupled to gravity, but on a degenerate
gravitational sector. Thereby the electromagnetic degrees of
freedom are decoupled and we can study them isolated from the complicated
gravitational dynamics. Note that a degeneracy of the metric does not
prevent electric and magnetic flows, and the densitized electric
and magnetic fields from being well-defined. The usual electromagnetic
Hamiltonian, however, vanishes, so we can study a static
electromagnetism only.

The plan of the paper is as follows: In Section 2 we will prove the
central assertion of this paper, namely the equivalence of the Abelian
$BF$-theory with the spherically symmetric sector of electromagnetism,
discuss the boundary conditions and present the reduced phase space
quantization. The corresponding spin network quantization, which is
also useful for a general $BF$-theory and not just for the special
case related to spherically symmetric electromagnetism, is developed
in Section 3 leading to the same results as in the reduced phase
space quantization of Section 2. Finally, we will recall results
obtained in the general symmetry reduction scheme of Ref.~\cite{SymmRed}
and compare them with the approaches of the present paper.  In the
appendices we will describe the dimensions used here, and recall the
classical reduction from Ref.~\cite{SphKlEM} as well as basics about
$U(1)$-spin networks.

\section{{\boldmath$BF$}-Theory}

As already noted above, the quantization of spherically symmetric
electromagnetism presented here is related to a $BF$-theory which
requires a vanishing magnetic field $\mu^a$.  We thus are lead to
study an Abelian $BF$-theory \cite{Horo:BF,BlauThom:BF} which has,
besides the Gau\ss\ constraint analogous to that of electromagnetism,
a second constraint $F_{ab}=\epsilon_{abc}\mu^c\approx0$ which
constrains the curvature of a $U(1)$-connection to vanish.

\subsection{Action and Constraints}

The theory we start with has as variables a two-form $B$ and a
$U(1)$-connection $\omega$ on a four-dimensional spacetime manifold of
the form $M=\Sigma\times\R$. The curvature of $\omega$ is given by
$F=\md\omega$ and appears in the $BF$-action (for the definitions of our
dimensions and the constants $q$ and $\alpha$ see
Appendix~\ref{a:units})
\begin{equation}\label{BFaction}
 S_{BF}=\frac{q}{\alpha}\int_MB\wedge\md\omega.
\end{equation}
We now insert the $3+1$-decomposition $B=\frac{1}{2}B_{ab}\md
x^a\wedge\md x^b+ B_{0a}\md t\wedge\md x^a$ and $\omega=\omega_0\md
t+\omega_a\md x^a$ to obtain the Hamiltonian formulation (a dot
denotes a time derivative)
\begin{eqnarray}\label{BFactionelec}
 S_{BF} & = & \frac{1}{2}\frac{q}{\alpha}\int_{\Sigma\times\R}
 \md^3x\md t\epsilon^{abc}\left(B_{ab}\dot{\omega}_c-B_{ab}\partial_c\omega_0
  +B_{0a}F_{bc}\right)\nonumber\\
 & = & \frac{q}{\alpha}\int_{\Sigma\times\R}\md^3x\md t\left(\epsilon^c
   \dot{\omega}_c-
  \epsilon^c\partial_c\omega_0+\frac{1}{2}\epsilon^{abc}\xi_aF_{bc}\right).
\end{eqnarray}
In the last step we introduced the field
$\epsilon^a:=\frac{1}{2}\epsilon^{abc}B_{bc}$, which will later be
identified with the electric field. These field components are
canonically conjugate to $\omega_a$ with the Poisson structure given
by Eq.~(\ref{Poisson}). The remaining components $\xi_a:=B_{0a}$
of $B$ and $\omega_0$ are Lagrange multipliers which upon variation
lead to the constraints
\begin{eqnarray}
 G[\omega_0]:=\int_{\Sigma}\md^3x\omega_0\partial_a\epsilon^a &\approx & 0,\\
 {\cal F}_0[\xi_a]:=\frac{1}{2}\int_{\Sigma}\md^3x\xi_a
 \epsilon^{abc}F_{bc} &\approx & 0.
\end{eqnarray}
Boundary conditions, which are important because we had to integrate
by parts, are discussed in the next subsection.

The Gau\ss\ constraint generates gauge transformations of $\omega$ as in
electromagnetism, whereas the new constraints ${\cal F}_0$ constrains the
magnetic field to vanish. More important here is the fact that ${\cal
  F}_0$ generates gauge transformations of $\epsilon^a$ which effect the
symmetry reduction. This is stated as
\begin{lemma}\label{BFrotsymmE}
  Let $\Sigma$ be a three-dimensional manifold carrying an action of
  the group $SO(3)$, and $(x,\vartheta,\varphi)$ be a (local) system
  of polar coordinates adapted to the spherical symmetry.

  The set of all spherically symmetric (time-dependent) fields
  $\epsilon^a=(\epsilon(t,x),0,0)$ with the boundary condition
  $\epsilon(t,\infty)=0$ is a set of representatives of the gauge
  equivalence classes of `electric' fields in the $BF$-theory
  Eq.~(\ref{BFaction}) vanishing at infinity.
\end{lemma}
\begin{proof}
  Because of $\{{\cal F}_0[\xi_a],\omega_b\}=0, \{{\cal
    F}_0[\xi_d],\epsilon^a\}= \alpha q^{-1}\epsilon^{abc}\partial_b\xi_c$ the
  gauge transformations generated by ${\cal F}_0$ lead to the addition
  of an exact two-form to the dual two-form $\epsilon_{abc}\epsilon^c$
  of $\epsilon^a$.
 
  First we show that two different symmetric electric fields
  $\epsilon_1^a$ and $\epsilon_2^a$ cannot lie in the same ${\cal
    F}_0$-gauge class. By assumption the difference
  $\delta\epsilon^a=\epsilon_1^a-\epsilon_2^a$ of these fields
  fulfills $\delta\epsilon^{\kt}= \delta\epsilon^{\kp}=0$ ($x$,
  $\vartheta$ and $\varphi$ are spherical coordinates). If that
  difference was a gauge transformation generated by some $\xi_a$,
  this function had to obey the equations
  $$
   \delta\epsilon^{\kt}=\partial_{\kp}\xi_x-\partial_x\xi_{\kp}=0\quad,\quad
   \delta\epsilon^{\kp}=\partial_x\xi_{\kt}-\partial_{\kt}\xi_x=0.
  $$
  This, in turn, would imply
  $$
   \partial_x\delta\epsilon^x=\partial_x(\partial_{\kt}\xi_{\kp}- 
   \partial_{\kp}\xi_{\kt})=
   \partial_{\kt}\partial_{\kp}\xi_x-\partial_{\kp}\partial_{\kt}\xi_x=0,
  $$
  i.e., the difference of the electric fields would be a constant
  which had to vanish due to the boundary condition. Therefore, each
  gauge class contains at most one spherically symmetric electric
  field.
  
  We now prove that each class contains at least one spherically
  symmetric electric field. In order to show this we need a vector field
  $N^a$ on $\Sigma$ which is spherically symmetric, i.e.,
  $N^a=(N(t,x),0,0)$, and which is subject to the conditions
  $\partial_aN^a=0$ and $\int_{S_x}\md^2S_aN^a=1$ for all $SO(3)$-orbits
  $S_x$ in $\Sigma$.  Such a field exists because, for the symmetry
  reduced theory to be non-trivial, we have to assume that there is at
  least one spherically symmetric electric field $\epsilon_0^a$, by
  means of which we can construct
  $$
     N^a(t,x):=\left(\int_{S_x}\md^2S_a\epsilon_0^a\right)^{-1}\epsilon^a_0.
  $$ 
  The existence of such a non-trivial field $\epsilon^a_0$ depends
  on the topology of $\Sigma$. It always exists in the manifolds of
  Appendix~\ref{a:reduct}.  The properties of $N^a$ postulated above follow
  from the ones of $\epsilon_0^a$ and the fact that
  $\int_{S_x}\md^2S_a\epsilon_0^a$ does not depend on $x$ (due to
  $\partial_a\epsilon^a_0=0$).

  Let $\epsilon^a(t,x,\kt,\kp)$ now be a field vanishing at infinity.
  Due to the properties of $N^a$ the averaged field
  $\overline{\epsilon}^a(t,x):= N^a\int_{S_x}\md^2S_b\epsilon^b$ is
  spherically symmetric and fulfills the Gau\ss\ constraint
  $\partial_a\overline{\epsilon}^a=0$. Furthermore, we have
  $\int_{S_x}\md^2S_a(\epsilon-\overline{\epsilon})^a=0$, which
  remains valid after replacing $S_x$ by an arbitrary closed surface.
  According to de Rham duality of homology and cohomology groups the
  difference of the two fields is cohomologically trivial and,
  therefore, exact:
  $\epsilon^a-\overline{\epsilon}^a=\epsilon^{abc}\partial_b\xi_c$ with
  an appropriate $\xi_c$. An electric field and its spherically
  symmetric average, therefore, lie in the same gauge class.

  Summarizing, we have proved that each gauge class contains exactly
  one spherically symmetric electric field.
\end{proof}

The meaning of this lemma is that the spherical symmetry reduction of
electromagnetism in its magnetically uncharged sector can be viewed as
gauge fixing of the transformations generated by the constraint ${\cal
  F}_0$ of the associated Abelian $BF$-theory. The remaining Gau\ss\ 
constraint has the same meaning in both theories generating the gauge
transformations $\omega_a\mapsto\omega_a+\partial_a\omega_0$.  Using
$F_{ab}=0$ and a fixed basis $([\omega^{(k)}])$ of $H^1(\Sigma,\R)$
any connection can be written as $\omega_a=\omega_a^{(k)}+\partial_a
l$ with some function $l\colon\Sigma\to\R$.  Each function $l$ can be
gauged to the spherically symmetric value $l=0$ by the gauge
transformation $l\mapsto l+\omega_0$ with $\omega_0:=-l$. Analogously,
$B_{ab}\md x^a\wedge\md x^b$ is closed due to the Gau\ss\ constraint
$G$, and ${\cal F}_0$ generates an additional exact two-form added to
$B_{ab}\md x^a\wedge\md x^b$.  This shows that on a manifold without
boundary the reduced phase space of $BF$-theory is given by the
product $H^1(\Sigma,\R)\times H^2(\Sigma,\R)$ of first and second de
Rham cohomology groups \cite{Horo:BF}.

However, in the manifolds used in the spherically symmetric context
(Appendix~\ref{a:reduct}) we have $H^1(\Sigma,\R)=0$, whereas
$H^2(\Sigma,\R)$ does not need to be even dimensional and is,
therefore, inappropriate as phase space. This is possible because we
use manifolds with boundary where $H^1(\Sigma,\R)=0$ and
$H^2(\Sigma,\R)$ do not have necessarily the same dimension.
Furthermore, the constraints are affected by the presence of a
boundary and the consideration of the preceding paragraph cannot be
applied unaltered: The function $l$ can now be gauged to be zero only
in the interior of $\Sigma$, whereas it remains arbitrary at the
boundary. Taking the boundary properly into account will thus lead to
new boundary degrees of freedom, which will render the reduced phase
space even dimensional.

\subsection{Surface Terms and Boundary Degrees of Freedom}

Before discussing boundary conditions we will slightly generalize in
the $BF$-theory context the manifolds defined in
Appendix~\ref{a:reduct} by increasing the number of boundary
components. Because we are interested mainly in the boundary degrees
of freedom, we will confine ourselves to manifolds with a trivial first
homology group only.

Besides the wormhole manifold $W^3:=\R\times S^2$ with $H_2(W^3,\Z)=\Z$
and boundary
$$
 \partial W^3=:\partial_{\infty}W^3=:\partial_+W^3\cup\partial_
  -W^3\cong S^2\:\dot{\cup}\:S^2
$$
with two boundary components at positive and negative infinity (the
boundary is a disjoint union of two $S^2$), we will use the punctured
manifolds $P_n^3:=\R^3\backslash\{p_1,\dots,p_n\}$ with
$H_2(P_n^3,\Z)=\Z^n$.  Equivalently, we can cut out of $\R^3$ a small
ball centered in each of the points $p_i$ resulting in new boundary
components $\partial_iP_n^3\cong S^2$, the full boundary
$$
 \partial
 P_n^3=\partial_{\infty}P_n^3\:\dot{\cup}\:\dot{\bigcup}_{i=1}^n
 \partial_iP_n^3
$$
having $n+1$ components. Similar to the wormhole manifold above we
denote with $\partial_{\infty}P_n^3$ the part of the boundary lying at
infinity, which topologically is just a specification of a
distinguished boundary component. Due to the non-trivial second
homology groups these manifolds allow topological electric charge, and
therefore we do not have to couple matter fields as sources of charge.
Of course, $W^3$ is homeomorphic to $P_1^3$, but the interpretation is
different as a spacelike section in the Reissner-Nordstr\o m manifold
as opposed to a point charge sitting in the origin.

For the constraints to be functionally differentiable we have to
impose boundary conditions, and to correct the action by boundary
terms. Boundary conditions for $BF$-theories have already been
discussed in Refs.~\cite{Wu:BFRand,Boundary,Vertex,BFSU}, but here we
choose different ones adapted to the interpretation as spherically
symmetric electromagnetism.

The variation of the constraints is
\begin{eqnarray}
 \delta G[\omega_0] & = &
  \int_{\Sigma}\md^3x\omega_0\partial_a\delta\epsilon^a=-\int_{\Sigma}\md^3x
  \delta\epsilon^a\partial_a\omega_0+\int_{\partial\Sigma}\md^2S_a\omega_0
  \delta\epsilon^a,\\
 \delta{\cal F}_0[\xi_a] & = & \int_{\Sigma}\md^3x\xi_a
  \epsilon^{abc} \partial_b\delta\omega_c =-\int_{\Sigma}\md^3x\epsilon^{abc}
  (\partial_b\xi_a)\delta\omega_c+\int_{\partial\Sigma}\md^2S_b
  \epsilon^{abc}\xi_a\delta\omega_c. \label{Fzero}
\end{eqnarray}
In order to achieve functional differentiability the surface integrals
have to vanish or to be compensated by appropriate boundary terms in
the action.

This can be enforced, first for the variation of $G$, by the condition
$\omega_0|_{\partial\Sigma}=0$ for gauge transformations. If we have
instead $\omega_0=O(1)$ on $\partial\Sigma$, the generated
transformation is viewed as symmetry transformation. For the surface
integral to vanish in this case we must require
$\delta\epsilon=O(r^{-(2+\delta)})$, $\delta>0$ on
$\partial_{\infty}\Sigma$ and $\delta\epsilon|_{\partial_i\Sigma}=0$
which is also necessary for symmetry transformations not to change the
charge. Surface variables are given by
\begin{equation}
 {\cal O}[\omega_0]:=\int_{\Sigma}\md^3x\epsilon^a\partial_a\omega_0=
 \int_{\partial\Sigma}\md^2S_a\omega_0\epsilon^a\quad,\quad\omega_0=O(1),
\end{equation}
further constrained by ${\cal F}_0$, however.

According to Lemma \ref{BFrotsymmE} the transformations generated by
${\cal F}_0$ are necessary for a symmetry reduction. Therefore, we
want to regard them as gauge transformations in any case without
specifying further boundary conditions on $\xi_a$. We need a
surface term in the action (\ref{BFactionelec}) with a variation
eliminating the surface integral in Eq.~(\ref{Fzero}). The corrected
action is
\begin{equation}
 S:=S_{BF}-\frac{q}{\alpha}\int_{\partial\Sigma\times\R}\md^2S_b\md
 t\epsilon^{abc} \xi_a\omega_c,
\end{equation}
leading to the Hamiltonian
\begin{eqnarray}\label{BFHam}
 H & = & \frac{q}{\alpha}\int_{\Sigma}\md^3x\left(\epsilon^a
  \partial_a\omega_0-
  \frac{1}{2}\epsilon^{abc}\xi_aF_{bc}\right)+\frac{q}{\alpha}
  \int_{\partial\Sigma}\md^2S_b \epsilon^{abc}\xi_a\omega_c\nonumber\\
 & = & -\frac{q}{\alpha}\int_{\Sigma}\md^3x\left(\omega_0
  \partial_a\epsilon^a+
  \frac{1}{2}\epsilon^{abc}\xi_aF_{bc}\right)+\frac{q}{\alpha}
  \int_{\partial\Sigma}\md^2S_b
  (\omega_0\epsilon^b+\epsilon^{abc}\xi_a\omega_c).
\end{eqnarray}
The boundary values of $\omega_0$ on $\partial\Sigma$ are prescribed
functions, which are determined by an external observer, depending on
the time variable $t$. In contrast, $\xi_a$ is regarded as
Lagrange multiplier also at the boundary leading to the corrected
curvature constraint
$$
  {\cal F}[\xi_a]={\cal F}_0-\int_{\partial\Sigma}\md^2S_b
  \epsilon^{abc}\xi_a\omega_c.
$$
Note that we did not specify boundary conditions for $\xi_a$ in the
context of ${\cal F}$. Therefore, variation of the boundary values
leads to the so-called natural boundary conditions. Thereby we obtain
the surface constraints
$n_a\epsilon^{abc}\omega_c|_{\partial\Sigma}\approx0$ ($n_a$ being the
normal on $\partial\Sigma$), which yield that $l$ is locally constant
on the boundary, i.e., constant on each boundary component, after
inserting $\omega_c=\partial_c l$. Together with
Lemma~\ref{BFrotsymmE} we can now see full equivalence to spherically
symmetric electromagnetism of Appendix \ref{a:reduct} (for the
manifolds $W^3$ or $P_1^3$):
\begin{theo}\label{BFrotsymm}
  The partially reduced phase space of the Abelian $BF$-theory
  obtained after solving only the constraint ${\cal F}$ is equivalent
  to the phase space of spherically symmetric electromagnetism.
\end{theo}

Before reducing the theory completely we check the algebra of
constraints.  Because $G$ and ${\cal F}$ contain either $\epsilon^a$
or $\omega_a$, we have
\begin{equation}
 \{G[\omega_0],G[\omega_0']\}= \{{\cal
   F}[\xi_a],{\cal F}[\xi_b']\}=0. 
\end{equation}
The mixed Poisson bracket is
\begin{eqnarray}
 \{G[\omega_0],{\cal F}[\xi_a]\} & = & -\frac{\alpha}{q}
  \int_{\Sigma}\md^3x
  (\partial_c\omega_0)\partial_b\xi_a\epsilon^{abc}=
  \frac{\alpha}{q}\int_{\Sigma}\md\omega_0\wedge\md\xi\nonumber\\
 & = & -\frac{\alpha}{q}\int_{\Sigma}\md(\md\omega_0\wedge\xi)=
  -\frac{\alpha}{q}\int_{\partial\Sigma}\md\omega_0\wedge\xi,
\end{eqnarray}
which vanishes for gauge transformations because then $\omega_0$ has
to vanish on the boundary. Therefore, the constraints are first class.
For the Poisson bracket to vanish we must have
$\md\omega_0|_{\partial\Sigma}=0$ which is also fulfilled for some special
symmetry transformations (for which
$\omega_0|_{\partial\Sigma}\not=0$, but locally constant). Because
$\xi$ is arbitrary at the boundary the surface variables ${\cal
  O}[\omega_0]$ are observables exactly if $\omega_0|_{\partial\Sigma}$ is
locally constant:
\begin{equation}
 \{{\cal O}[\omega_0],{\cal F}[\xi_a]\}=0.
\end{equation}

The observables ${\cal O}[\omega_0]$ (with unrestricted $\omega_0$)
have already appeared in Ref.~\cite{Boundary}, together with
additional surface observables which are integrals of $\omega_a$
associated with boundary values of $\xi_a$. These latter
observables are excluded here by our special boundary conditions (free
boundary values of the Lagrange multiplier $\xi_a$). It also leads
to the restriction of $\omega_0$ in ${\cal O}[\omega_0]$ to be locally
constant on $\partial\Sigma$. The special treatment of $\xi_a$,
leading to these two effects, is crucial for the identification with
the spherically symmetric sector of electromagnetism (see also
Eq.~(\ref{BoundConst}) below); and we will see that the remaining
surface observables are just the correct ones for this application.

\subsection{Reduction and Quantization}

The constraints are easy to solve: ${\cal F}$ forces the connection
$\omega_a$ to be flat, i.e., $\omega_a=\partial_a l$ for some
$l\colon\Sigma\to\R$. Since $G$ generates the gauge transformation
$l\mapsto l+\omega_0$ with an arbitrary $\omega_0$ vanishing on the
boundary, only the boundary values of $l$ have physical
meaning. Furthermore, due to the boundary constraints
\begin{equation}\label{BoundConst}
 {\cal C}[\xi_b|_{\partial\Sigma}]:=\int_{\partial\Sigma}\md
 S_a\epsilon^{abc}\xi_b\omega_c=\int_{\partial\Sigma}\md
 S_a\epsilon^{abc}\xi_b\partial_cl\approx0
\end{equation}
$l$ has to be constant on each boundary component because the
$\xi_a$ are arbitrary at $\partial\Sigma$. This is the most important
consequence of our special boundary conditions introduced above.

The physical degrees of freedom associated with $\epsilon^a$ can also be
localized at the boundary and given by integrals
\begin{equation}
 p^A:=\int_{\partial_A\Sigma}\md^2S_a\epsilon^a
\end{equation}
over each of the $n+1$ boundary components, i.e.\ $p^A={\cal
  O}[\omega^A_0]$ with $\omega^A_0|_{\partial_B\Sigma}=\delta^A_B$.
They are not all independent, however, because of
$\sum_Ap^A=\int_{\partial\Sigma}\md^2S_a\epsilon^a=
\int_{\Sigma}\md^3x\partial_a\epsilon^a=0$ as a consequence of the
Gau\ss\ constraint.  As above, the constraints imply that the class of
$\epsilon_{abc}\epsilon^c$ in the second de Rham cohomology group
represents the physical degree of freedom specified by its evaluation
on all classes of the second homology group. Choosing as
representatives for a basis of the second homology groups of the two
manifolds $(\partial_+\Sigma)$ and $(\partial_i\Sigma)_{1\leq i\leq
  n}$, respectively, we arrive at the independent observables $p^+$
and $p^1,\ldots,p^n$.

From Eq.~(\ref{BFHam}) we derive the reduced Hamiltonian: At first
we insert the bulk constraints to obtain 
\begin{equation}
 H'=\frac{q}{\alpha}\int_{\partial\Sigma}\md^2S_a
   (\omega_0\epsilon^a-\epsilon^{abc}\xi_b\omega_c)
   =\frac{q}{\alpha}\int_{\partial\Sigma}\md^2S_a(\dot{l}\epsilon^a-
   \epsilon^{abc}\xi_b\partial_cl).
\end{equation}
In the second part of this equation we provided a time dependence for
$l$ by defining $\dot{l}:=\omega_0$, which formally extends the
relation $\omega_a=\partial_a l$ to the four-dimensional connection on
$\Sigma\times\R$.

The boundary values of $\xi_a$ are the remaining Lagrange
multipliers, and their variation leads to the boundary constraints
${\cal C}$ of Eq.~(\ref{BoundConst}). The solution of these requires a
locally constant $l$ on the boundary eliminating the second term in
$H'$:
\begin{equation}
 H''=\frac{q}{\alpha}\int_{\partial\Sigma}\md^2S_a\dot{l}\epsilon^a
   =\frac{q}{\alpha}\sum_A\dot{l}_A\int_{\partial_A\Sigma}\md^2S_a\epsilon^a
   =\frac{q}{\alpha}\sum_A\dot{l}_Ap^A,
\end{equation}
where $A$ runs over all $n+1$ boundary components and $l_A$ is the
constant value of $l$ on the component $\partial_A\Sigma$.

Up to now the constraints are not solved completely: There remains the
condition $\sum_Ap^A=\int_{\partial\Sigma}\md^2S_a\epsilon^a\approx0$
implied by $G\approx 0$. Therefore, only $n$ of the $n+1$ boundary
variables $p^A$ are independent. At the same time, the remaining
constraint generates the gauge transformation $l_A\mapsto l_A+c$ with
some $c$ being constant on the full boundary (not just locally
constant). One of the $l_A$ can thereby be gauged to zero, and we end
up with only $n$ independent values of the $l_A$. In our manifolds we
will choose the gauge fixing $l_-=0$ in $W^3$ and $l_{\infty}=0$ in
$P_n^3$, respectively. Finally we obtain the reduced Hamiltonian
\begin{equation}\label{Hred}
 H_{red}=\frac{q}{\alpha}\sum_A\dot{l}_Ap^A\quad,\quad A =+\mbox{
   or }A\in\{1,\dots,n\}.
\end{equation}
A comparison with Appendix~\ref{a:reduct} shows that this is, on the
manifolds $W^3$ or $P_1^3$, the reduced Hamiltonian of spherically
symmetric electromagnetism with the prescribed function of $t$ being
$\dot{l}=U$, which reveals that its boundary dynamics is equivalent to
that of $BF$-theory, too. The canonical variables
$$
 (p^A,q\alpha^{-1}\Phi_A)_{A=+\mbox{ \footnotesize or }A\in\{1,\dots,n\}}
$$
are action-angle coordinates of the reduced Hamiltonian. The
equations of motion are solved by
\begin{eqnarray*}
 p^A & = & c_A\\
 \Phi_A & = & c'_A-l_A(t)
\end{eqnarray*}
with constants $c_A,c'_A$ to be specified by initial values. 

This system with phase space $T^*\R^n$ can be
quantized without problems. As Hilbert space we choose
$L^2(\R^n,\md^nx)$, $n+1$ being the number of boundary components. In
the $\Phi$-representation states are given by
$\psi(\Phi_1,\dots,\Phi_n)$, acted on by $\hat{\Phi}_A$ and
$\hat{p}^A$ as usually:
\begin{equation}
 \hat{\Phi}_A\psi=\Phi_A\psi\quad,\quad\hat{p}^A\psi
 =\frac{\hbar}{i}\frac{\alpha}{q}\pd{\Phi_A}\psi=\frac{q}{i}\pd{\Phi_A}\psi.
\end{equation}
This quantization leads to a continuous spectrum of the charges $p^A$,
but a quantization condition can be imposed by $\Phi_A\in
S^1\cong\R/2\pi\Z$, justified by the fact that $\Phi$ represents a
${\cal L}U(1)$ element: $l$ and $l+2\pi$ yield the same element $\exp
il=\exp i(l+2\pi)$ of the gauge group. This periodical identification
of the phase space leads to charge quantization: Simultaneous
eigenstates of the $\hat{p}^A=-iq\pd{\Phi_A}$ are given by
$\psi_{\{K_A\}}(\Phi_1,\ldots,\Phi_n)=\prod_A\exp iK_A\Phi_A$ with
eigenvalue $qK_A$ of $\hat{p}^A$. The periodic identification demands
$K_A\in\Z$ leading to
\begin{equation}\label{spectrum}
 p^A\in q\Z\quad\mbox{ for all $A$}
\end{equation}
which is the observed charge quantization with a `fundamental charge'
$q$. Its value, however, cannot be determined because the theory
contains one free parameter $\alpha$. In the present context the
periodic identification of $\Phi_A$ looks somewhat ad-hoc, but the
quantization condition (\ref{spectrum}) will arise in the following
spin network quantization more naturally.

\section{Spin Network Quantization of {\boldmath$BF$}-Theory}

In the preceding section we arrived at a quantization of Abelian
$BF$-theory, which can be interpreted as a quantum theory of
spherically symmetric electromagnetism. In order to compare with the
results of the quantum symmetry reduction procedure it is, however,
more instructive to present a spin network quantization, too. The
solution of the constraints will be given in the same order as in the
symplectic reduction of the previous section: We first solve the Gau\ss\
constraint, then the curvature constraint to arrive at a boundary
theory, and finally the boundary constraints. Our notation for
$U(1)$-spin networks, which are extensively used in this section, is
described in Appendix~\ref{a:U1spin}.

\subsection{Gau\ss\ Constraint}

Of course, the Gau\ss\ constraint can be solved by using only gauge
invariant spin networks, i.e., those with $k_v=0$ for each vertex
$v$, but for the sake of completeness we will give also a quantization
of the classical constraint $G$.

Let $\gamma$ be a graph and $f_{\gamma}$ be a cylindrical function
which depends on a connection $\omega_a$ only via the edge
holonomies $\eta_e$ for all $e\in E(\gamma)$. On that function the
Gau\ss\ constraint acts as
\begin{eqnarray*}
 \hat{G}[\omega_0]f_{\gamma} & = & \int_{\Sigma}\md^3x\omega_0
  \partial_a\hat{\epsilon}^af_{\gamma}=\frac{\hbar}{i}\frac{\alpha}{q}
  \int_{\Sigma}\md^3x\omega_0(x)\partial_a\fd{\omega_a(x)}f_{\gamma}\\
 & = & q\sum_{e\in E(\gamma)}\int_{\Sigma}\md^3x\omega_0(x) \int_e\md
  t\dot{e}^a\partial_a\delta(x,e(t))\eta_e\pd{\eta_e}f_{\gamma}\\
 & = & -q\sum_{e\in E(\gamma)}\int_{\Sigma}\md^3x\omega_0(x) \int_e\md
  t\td{t}\delta(x,e(t))\eta_e\pd{\eta_e}f_{\gamma}\\
 & = & -q\sum_{e\in E(\gamma)}\int_{\Sigma}\md^3x\omega_0(x)
  (\delta(x,e(1))-\delta(x,e(0)))\eta_e\pd{\eta_e}f_{\gamma}\\
 & = & -q\sum_{e\in E(\gamma)}(\omega_0(e(1))-\omega_0(e(0)))
 \eta_e\pd{\eta_e}f_{\gamma} =-q\sum_{v\in V(\gamma)}\omega_0(v)
 \sum_{e\in E(\gamma)}\sgn(v,e)\kappa_ef_{\gamma}.
\end{eqnarray*}
Recall the definition of $\sgn(v,e)$ given in Appendix~\ref{a:U1spin}
which implies that in the last sum only edges incident in $v$
contribute.

The $\hat{G}[\omega_0]$ commute with one another which means that the
classical algebra of constraints is represented anomaly-free.  Applied
to a spin network state $T_{\gamma k}$ the constraint yields
$\hat{G}[\omega_0]T_{\gamma k}= -q\sum_{v\in
  V(\gamma)}\omega_0(v)k_vT_{\gamma k}$, implying that the solution
space of the constraint is given by gauge invariant spin networks with
$k_v=0$ for all vertices $v\not\in\partial\Sigma$, as anticipated.
This constrains, however, only vertices in the interior of $\Sigma$
because at the boundary we had to demand
$\omega_0|_{\partial\Sigma}=0$ in the constraint. Therefore, $k_v$ for
$v\in\partial\Sigma\cap V(\gamma)$ is arbitrary meaning that at the
boundary edges of a spin network can end. This is analogous to the
boundary observables ${\cal O}$ appearing in the second section, and
it makes possible electric charge. The electric charge $Q[S]$ enclosed
by a closed surface $S\subset\Sigma$ is given by the integral
$\int_S\md^2S_a\epsilon^a$ in the classical case, which depends only
on the homology class of $S$. This lead us to use topologies of
$\Sigma$ with non-trivial second homology groups to allow electric
charge, and it follows classically from the Stokes theorem in a
well-known fashion: If $S_1$ and $S_2$ are the equally oriented
boundary components of a domain $B\subset\Sigma$, we have
$0=\int_B\md^3x\partial_a\epsilon^a=\int_{S_1}\md^2S_a\epsilon^a
-\int_{S_2}\md^2S_a\epsilon^a$ as a consequence of the Gau\ss\ 
constraint $\partial_a\epsilon^a\approx 0$.

The reason for our dwelling on that point is that in the quantum
theory there is an analogous, but quite differently, namely
topologically realized version. Here, the Gau\ss\ constraint manifests
itself in the condition $k_v=0$. At first we quantize the charge
functional applied to a function cylindrical with respect to a graph
$\gamma$ which is chosen such that all the intersection points $v\in
V(\gamma\cap S)$ are vertices of $\gamma$:
\begin{eqnarray}
 \hat{Q}[S] & = & \int_S\md^2yn_a\hat{\epsilon}^a
  =q\int_S\md^2y\sum_{e\in E(\gamma)}\int_e\md tn_a\dot{e}^a
  \delta(y,e(t))\eta_e\pd{\eta_e}\nonumber\\
 & = & q\sum_{v\in V(\gamma)\cap S}\sum_{e\ni v}\sgn(e,S)
  \eta_e\pd{\eta_e}.\label{charge} 
\end{eqnarray}
Here, $\sgn(e,S)$ is the intersection number of $e$ with $S$ which is
defined to be $\frac{1}{2}$ if $e\cap S\subset\partial e$. The charge
of a spin network state $T_{\gamma k}$ is proportional to $\sum_{e\cap
  S\not=\emptyset}\sgn(e,S)k_e$ which can be interpreted as the
intersection number of $S$ with a curve associated to $T_{\gamma k}$.
This curve, which is disconnected in general, can be constructed by
stacking $|k_e|$ copies of each edge $e\in E(\gamma)$ on top of each
other. All such copies incident in a vertex $v$ can be linked there to
form curves with no endpoints in $v$ if and only if $k_v=0$. In the
interior of a domain $B$ as above we therefore obtain pieces of curves
ending only at the boundary $\partial B=S_1\cup S_2^{\star}$
($S^{\star}$ is $S$ in opposite orientation) if and only if the spin
network state is gauge invariant. If there are only divalent vertices
at $\partial B$, the charges $\hat{Q}[S_1]T_{\gamma,k}$ and
$\hat{Q}[S_2]T_{\gamma,k}$ are equal being given by intersection
numbers of homologically equivalent closed surfaces with a closed
curve: Each curve entering $B$ through $S_1$ has to leave $B$ either
again through $S_1$, which does not contribute to both charges
measured by $S_1$ and $S_2$, or it runs through $S_2$ contributing to
the two charges the same amount.

\subsection{Curvature Constraint}

As opposed to the Gau\ss\ constraint the curvature constraint cannot
be solved in a subspace of the space of cylindrical functions, but it
has to be solved by means of a rigging map \cite{ALMMT}. This map can
be written formally as multiplication with a delta function supported
on the space of flat connections which will be constructed in this
subsection.

\subsubsection{The Space \boldmath$\Anb$ of Pure Gauge Connections}

In the simply connected topologies used here $F=0$ means that
$\omega_a=\partial_a l$ is pure gauge. Holonomies associated with an
edge $e$ are
$$
 \eta_e(l)=\exp\left(i\int_e\md l\right)=\exp
  \Bigl(i(l(e(1))-l(e(0)))\Bigr),
$$
which depend on $l$ only in the starting point $e(0)$ and the end
point $e(1)$ of $e$. A spin network state, therefore, depends only on
the values of the gauge potential $l$ in its vertices:
$$
 \prod_{e}\exp(i\:\sgn(v,e)k_el(v))=\exp(ik_vl(v)).
$$ 
By using the pure gauge connection each point $v\in\Sigma$ is mapped
to a $U(1)$ group element $\lambda(v):=\exp(il(v))$, and a spin
network state evaluated in this connection takes the form $\prod_{v\in
  V(\gamma)}\lambda(v)^{k_v}$. The function $\lambda\colon\Sigma\to
U(1)$ is smooth for a classical connection, but it will be generalized
to an arbitrary function in the course of quantization:
\begin{defi}
  $\An:=\{\lambda\colon\Sigma\to U(1)\mbox{ \rm smooth}\}$ is the space
  of {\em pure gauge connections}.

 The space of {\em generalized pure gauge connections} is
 $\Anb:=\{\lambda\colon\Sigma\to U(1)\}$.
\end{defi}
Analogously to $\Gb$ in Ref.~\cite{ALMMT}, the space $\Anb$ can be
constructed as a projective limit with index set being the set of all
finite subsets $\sigma\subset\Sigma,|\sigma|\in\N_0$ of $\Sigma$, and
with cylindrical spaces $\Ans:=U(1)^{\sigma}$ being the spaces of all
maps from $\sigma$ to $U(1)$ and projections
$p_{\sigma'\sigma}(\lambda_{\sigma})= \lambda_{\sigma}|_{\sigma'}$ for
$\lambda_{\sigma}\in\Ans$.  Then we have
$$
 \Anb=\projlim_{\sigma\subset\Sigma}\Ans=
 \projlim_{\sigma\subset\Sigma}U(1)^{\sigma} \equiv U(1)^{\Sigma}.
$$

A cylindrical basis of functions on $\Anb$ associated to the index set
of all finite subsets $\sigma$ together with labelings
$k\colon\sigma\to\Z\backslash\{0\}$ is given by the functions
$t_{\sigma,k}(\lambda) :=\prod_{p\in\sigma}\lambda(p)^{k_p}$. For
$\sigma$ fixed these are the monomials in the finitely many variables
$\lambda(p),p\in\sigma$, which certainly span the space of functions
on $\Ans$ modulo functions which are constant in some $\lambda(p)$,
i.e., functions on a space ${\cal A}_{0,\sigma'}$ with
$\sigma'\subset\sigma$. Analogously to the Ashtekar-Lewandowski
measure we can define a measure on $\Anb$ cylindrically:
\begin{lemma}\label{measure}
  A diffeomorphism invariant probability measure on $\Anb$ is given by
  $$
    \mu(f):=\int_{\Anb}\md\mu f:=\mu_{\sigma}(f_{\sigma})
    :=\int_{U(1)^{|\sigma|}}\md\mu_H^{|\sigma|}
    (\lambda_1,\dots,\lambda_{|\sigma|})f_{\sigma}
    (\lambda_1,\dots,\lambda_{|\sigma|})
  $$
  for some representative $f_{\sigma}$ of $f$. {\rm $\Diff(\Sigma)$}
  acts on $\Ans$ by $U(\phi)f_{\sigma}=f_{\phi(\sigma)}$.

  The monomials $t_{\sigma,k}$ form an orthonormal basis with respect
  to this measure.
\end{lemma}
\begin{proof}
 The cylindrical consistency condition for the measure and its
 normalization as well as diffeomorphism invariance follow from
 properties of the Haar measure.

 If $t_{\sigma,k}$ and $t_{\sigma',k'}$ are two monomials, then
 \begin{equation}\label{innprodmono}
  \ScProd{t_{\sigma,k}}{t_{\sigma',k'}}
  =\mu(\overline{t_{\sigma,k}}t_{\sigma',k'})
  =\int_{U(1)^{|\sigma\cup\sigma'|}}\md\mu_H^{|\sigma\cup\sigma'|}(\lambda)
  \prod_{p\in\sigma\cup\sigma'}\lambda(p)^{k'_p-k_p}
  =\delta_{\sigma\sigma'}\delta_{kk'}
 \end{equation}
 proving orthonormality.
\end{proof}

For $\sigma=V(\gamma)$ and $k$ being the vertex labeling of a spin
network state $T_{\gamma,k'}$ we have $t_{\sigma,k}(\lambda)
=T_{\gamma,k'}(\lambda^{-1}\md\lambda)$ showing that the
$t_{\sigma,k}$ emerge by restriction of spin network states to pure
gauge connections. To each spin network $T_{\gamma,k'}$ we can
associate a monomial $t_{\sigma,k}=:\partial T_{\gamma,k'}$ and
continue the operation $\partial$ to the space $\Phi$ of all
cylindrical functions on $\Ab$. Formally, we can write
$t_{\sigma,k}=\delta({\cal F})T_{\gamma,k'}$ with
$$
 \delta({\cal F}):=\prod_{e\subset\Sigma}\prod_{p\in\Sigma} 
 \int_{U(1)}\md\mu_H(\lambda(p))
 \delta\left(A_e,\lambda(e(0))^{-1}\lambda(e(1))\right).
 $$ Given a cylindrical function $f_{\gamma}$ we have to interpret
 $\delta({\cal F})f_{\gamma}$ as a distribution on the space of
 cylindrical functions, and equivalently $\delta({\cal
   F})\colon\Phi\to\Phi'$ as a rigging map according to
\begin{eqnarray}\label{deltaF}
 (\delta({\cal F})f_{\gamma})(g_{\gamma'}) & := & \int_{\Anb}\md\mu(\lambda)
  \prod_{e\in E(\gamma)\cup E(\gamma')}\int_{U(1)}\md\mu_H(A_e)
  \delta\left(A_e,\lambda(e(0))^{-1}\lambda(e(1))\right)\overline{f}_{\gamma}
  g_{\gamma'}\nonumber\\
  & = & \int_{\Anb}\md\mu(\lambda)\overline{\partial f}_{\gamma}
   \partial g_{\gamma'}.
\end{eqnarray}
This map solves the constraint ${\cal F}$ on a subspace of $\Phi'$.

\subsubsection{Boundary Spin Networks}
\label{s:boundary}

Up to now we solved the constraints $G$ and ${\cal F}$ separately. To
solve them together we have to investigate the space $\AnG$. In
contrast to $\An$, the space ${\cal G}$ consists of functions
$g\colon\Sigma\to U(1)$ which have to become unity on the boundary
$\partial\Sigma$, and it acts on $\Anb$ as
$(g\cdot\lambda)(p)=g(p)\lambda(p)$ (note that $\lambda$ is the
exponentiated gauge potential of a connection; therefore, $g$ does not
act by conjugation). For $\Sigma$ without boundary we have
$\AnG=\{1\}$, and similar to the boundary observables in the classical
theory non-trivial states emerge only in presence of a boundary. We
have the projective spaces
$$
 \Ans/{\cal G}_{\sigma}=\{\lambda\colon\sigma\to U(1)|\lambda(p)=1
 \mbox{ if }p\not\in\partial\Sigma\}
$$
with limit
$$
 \AnGb=\Anb/\overline{{\cal G}}=\{\lambda\colon\partial\Sigma\to U(1)\}.
$$
All degrees of freedom are localized at the boundary of $\Sigma$
motivating
\begin{defi}
  The space of functions on the space $\AnGb$ of gauge invariant pure
  gauge connections is spanned by {\em boundary spin networks}
  $t_{\sigma,k}$ with finite sets $\sigma\subset\partial\Sigma$ and
  labelings $k\colon\sigma\to\Z\backslash\{0\}$. The associated
  boundary state is given by
  $t_{\sigma,k}(\lambda):=\prod_{p\in\sigma}\lambda(p)^{k_p}$.
 
  These functions $t_{\sigma,k}$ span the {\em boundary Hilbert space}
  when completed with respect to the measure $\mu$ of
  Lemma~\ref{measure}.
\end{defi}

The fundamental operations can also be projected down from the
spin network basis by inserting pure gauge connections. The holonomy
to an edge $e$ in $\Sigma$ with $e\cap\partial\Sigma=e(1)=:p$ is
$$
 \eta_e=\frac{\lambda(p)}{\lambda(e(0))}.
$$
Gauge invariant is only $\lambda(p)$ leading to the multiplication
operator $\lambda_p:=\lambda(p)$ instead of $\eta_e$. Its action on a
boundary spin network $t_{\sigma,k}$ is to increase $k_p$ by $1$.

The other fundamental operator associated to $e$ is the derivative
operator $\kappa_e$:
$$
 \kappa_e t_{\sigma,k}=\eta_e\pd{\eta_e}t_{\sigma,k}
 =\frac{\lambda(p)}{\lambda(e(0))}\pd{\lambda(e(0))^{-1}\lambda(p)}
 t_{\sigma,k}=\lambda(p)\pd{\lambda(p)}t_{\sigma,k}=k_pt_{\sigma,k},
$$
where we used independence of $t_{\sigma,k}$ on $\lambda(e(0))$. This
operator also acts only in the point $p$ and can be written as
$$
 \kappa_p:=\lambda_p\pd{\lambda_p}.
$$

In this way we obtain multiplication and differentiation on the
monomials, out of which, together with the adjoint of $\lambda_p$, we
can build all local operators. The space of boundary spin networks
resembles the construction of a Fock space with `one-particle Hilbert
spaces' $L^2(U(1),\md\mu_H)$ associated to each point
$p\in\partial\Sigma$, but without any symmetrization (it is neither
bosonic nor fermionic). The operators $\lambda_p$ and $\kappa_p$ act
as creation and number operator, respectively. In contrast to a usual
Fock space (as, e.g., used in Ref.~\cite{Boundary} for similar
purposes) we can represent the full group of boundary diffeomorphisms
on our boundary Hilbert space with a diffeomorphism invariant measure,
which is, of course, a consequence of our usage of spin network
techniques.

\subsection{Boundary Constraints}

The bulk constraints $G$ and ${\cal F}$ are now solved on $\AnGb$.
However, the boundary constraints ${\cal C}$, which force $l$ to be
constant on each of the $n+1$ boundary components, still remain to be
solved. We have to impose them on boundary spin network states by
restricting these functions to locally constant $\lambda=\lambda_A$ on
each $\partial_A\Sigma$. A restricted state is completely determined
by an integer
$$
 K_A:=\sum_{p\in\sigma\cap\partial_A\Sigma}k_p
$$
for each boundary component $\partial_A\Sigma$, which can be seen from
the calculation
\begin{equation}\label{tsk}
 t_{\sigma,k}(\lambda)|_{{\cal C}=0}=\prod_A\lambda_A^{
   \sum_{p\in\sigma\cap\partial_A\Sigma}k_p}=\prod_A\lambda_A^{K_A}
  =:\ket{K_1,\ldots,K_{n+1}}. 
\end{equation}

As in the classical reduction there remains one last condition
following from gauge invariance. The boundary spin networks descend
from gauge invariant spin networks in the course of constraint
reduction, which implies $\sum_AK_A=0$. Again, only $n$ of the $n+1$
numbers $K_A$ are to be chosen freely. Accordingly, the $\lambda_A$
can be multiplied by some $\lambda_0$ which is constant on the whole
boundary, because states change then by multiplication with a factor
$\lambda_0^{\sum_AK_A}=1$. This freedom can be fixed by imposing the
condition $\lambda_A=1$ for some fixed boundary component $A$,
analogous to the classical case, and discarding its charge $K_A$.  The
states $\ket{K_1,\ldots,K_n}$ labeled by the remaining $n$ charges
build an orthonormal basis of the physical Hilbert space
$\Hphys=L^2(U(1)^n,\md\mu_H^n)$ with the inner product descending from
the space of boundary spin networks:
\begin{equation}\label{innprod}
 \DirB{K_1,\ldots,K_n}{K_1',\ldots,K_n'}:=
 \ScProd{t_{\sigma,k}}{\delta({\cal C})t_{\sigma',k'}}
 :=\prod_A\int_{U(1)}\md\mu_H(\lambda_A)
 \lambda_A^{K_A'-K_A}=\prod_A\delta_{K_A,K_A'}
\end{equation}
where, formally,
\begin{equation}\label{deltaC}
 \delta({\cal C}):=\prod_A\prod_{p\in\partial_A\Sigma} \int_{U(1)}\md\mu_H
 (\lambda_A)\delta(\lambda_p,\lambda_A).
\end{equation}

Finally, we need a representation of the Poisson algebra of the
canonical variables $(p^A,\Phi_A)$ on the physical Hilbert space. The
operators are to be built from the boundary operators $\lambda_p$ and
$\kappa_p$, and they can be deduced from their action on
three-dimensional spin network states.

According to Appendix~\ref{a:reduct}, $p^A$ and $\Phi_A$ are in
generalization from the spherically symmetric case, i.e., from the
manifolds $W^3$ or $P_1^3$, given by the charge on the boundary
component $\partial_A\Sigma$ and by the holonomy associated with a
radial curve $B_A$ ending on $\partial_A\Sigma$:
\begin{equation}
 p^A=\int_{\partial_A\Sigma}\md^2S_a\epsilon^a\quad,\quad\Phi_A
 =-\int_{B_A}\omega=i\log\eta_{B_A}.
\end{equation}
These expressions are the same as the reduced phase space variables of
the preceding section. To obtain independent variables we have to
choose $n$ out of the $n+1$ boundary components (as in
Eq.~(\ref{Hred}), for instance), the index $A$ running over them in
the following.  The excluded component can be used to provide a
starting point for the curves $B_A$ (for $\Phi_A$ to be gauge
invariant $B_A$ cannot start in the interior of $\Sigma$). In this
way, each curve intersects only one of the distinguished components
$\partial_A\Sigma$.

$p^A$ is quantized by using Eq.~(\ref{charge}). The surface $S$ in
this equation is chosen to be homologically equivalent to the boundary
component $\partial_A\Sigma$ and lying in the interior of $\Sigma$.
$S$ must not be the boundary component itself because this would
introduce a factor of $\frac{1}{2}$ since all edges would end on $S$.
Note that charges are defined in the classical calculation also by
choosing a surface in the interior and computing the limit where this
surface approaches the boundary at infinity (which is, of course, only
necessary if there is no Gau\ss\ law, as e.g. for the ADM mass in a
theory of gravity).  However, for a boundary around a point charge we
could equally well integrate over the boundary in the classical
theory. In quantum theory this is no longer the case due to the
distributional nature of generalized connections. On boundary spin
networks we obtain
$$
 \hat{p}^At_{\sigma,k}=q\sum_{p\in\sigma\cap\partial_A\Sigma}\lambda_p
 \pd{\lambda_p}t_{\sigma,k}=
  q\sum_{p\in\sigma\cap\partial_A\Sigma}k_pt_{\sigma,k}=qK_At_{\sigma,k},
$$
which is to be projected into the physical Hilbert space:
\begin{equation}\label{hatpA}
 \hat{p}^A\ket{K_1,\ldots,K_n}=qK_A\ket{K_1,\ldots,K_n}.
\end{equation}

On spin network states the basic multiplication operator is not
$\Phi_A$, but the holonomy
$$
 \eta_A:=\eta_{B_A}=\exp\left(i\int_{B_A}\omega\right)=\exp(-i\Phi_A)
$$
with operation
\begin{equation}\label{hatPhiA}
 \exp(-i\hat{\Phi}_A)\ket{K_1,\ldots,K_n}
 =\hat{\eta}_A\ket{K_1,\ldots,K_n}=\ket{K_1,\ldots,K_A+1,\ldots,K_n}
\end{equation}
because $\eta_A$ reduces to multiplication with $\lambda(B_A(1))$ (see
Subsection \ref{s:boundary}, $B_A(1)$ denotes the endpoint of $B_A$).
The operator $\hat{\eta}_A$ can be interpreted as shifting charge from
the excluded boundary component to the component $\partial_A\Sigma$
along the curve $B_A$ (or rather its diffeomorphism class). In this
way, the total charge situated on all the boundary components remains
constant, namely zero.

We can now quote from Ref.~\cite{SymmRed} the following

\begin{theo}
 The Equations (\ref{hatpA}) and (\ref{hatPhiA}) define a
 representation of the classical Poisson $\star$-algebra on {\rm $\Hphys$}.
\end{theo}
\begin{proof}
 The proof is the same as in Ref.~\cite{SymmRed} except for an obvious
 generalization to $n$ variables.
\end{proof}

We note that the adjointness relations --~$\hat{p}^A$ being
self-adjoint and $\hat{\eta}_A$ being unitary~-- uniquely (up to a
constant factor) determine the inner product (\ref{innprod}) which was
derived by descending from the Ashtekar-Lewandowski measure. Moreover,
holonomy variables of spin network quantization turn out to be well
suited to represent the classical algebra of observables. As opposed
to Ref.~\cite{Vertex} we did not have to use a normal ordering to define
charge creation (or rather shifting) operators: Due to the basic
assumption of every spin network quantization, namely that holonomies
are well defined in quantum theory, the operators $\hat{\eta}_A$ are
perfectly well defined in our Hilbert space.

In complete analogy to the reduced phase space method we arrived at
the same quantum theory with one degree of freedom per boundary
component given by the electric charge. Moreover, we obtain
automatically a discrete charge spectrum (this has already been
observed in Ref.~\cite{MaxwellSpinnet} in case of unreduced
$U(1)$-spin networks) with eigenvalues $qK_A$ of the charge operator
belonging to the boundary component $\partial_A\Sigma$ being integer
multiples of $q$, which is however undetermined. This leads again to
the charge spectrum (\ref{spectrum}).

\subsection{Rigging Map}

As noted already, the curvature and boundary constraints cannot be
solved in a subspace of the space $\Phi$ spanned by spin network
states, but they have according to refined algebraic quantization
\cite{ALMMT} to be solved in its topological dual $\Phi'$. What we
have to do now is to present a rigging map implementing the
constraints. This can be constructed by using partial rigging maps
corresponding to the curvature and boundary constraints, respectively.

The basic ingredient for the rigging map $\eta_1\colon\Phi\to\Phi'$
has already been given in Eq.~(\ref{deltaF}). There we named it more
pictorially $\eta_1 f_{\gamma}:=\delta({\cal F})f_{\gamma}$.

In a second step we have to implement the boundary constraint ${\cal
  C}$. With the help of $\eta_1$ we went to the space $\Psi$ of
boundary spin networks $t_{\sigma,k}$, which is interpreted as a
subspace of $\Phi'$ analogously to Eq.~(\ref{deltaF}). Now we have to
start from $\Psi$ to go over to its dual $\Psi'$. Again, this can
formally be done by means of a delta distribution, $\delta({\cal C})$
in Eq.~(\ref{deltaC}), which enforces $\lambda$ to take the constant
value $\lambda_A$ at each boundary component $A$. The result of a
multiplication with $\delta({\cal C})$ is a function depending
only on the $n$ variables $\lambda_A$ which we denoted above as
$\ket{K_1,\ldots,K_n}=\prod_A\lambda_A^{K_A}$ (see Eq.~(\ref{tsk})).
This function is in the dual of $\Psi$ by means of
\begin{eqnarray}\label{etab}
 & & \ket{K_1,\ldots,K_n}(t_{\sigma,k}):=\prod_A
  \delta_{K_A,\sum_{p\in\sigma\cap\partial_A\Sigma}k_p}\\
 & & \quad = \int_{U(1)^n}\md^n\mu_H(\lambda_1,\dots,\lambda_n)
  \prod_A\lambda_A^{\sum_{p\in\sigma\cap\partial_A\Sigma}k_p-K_A}\nonumber\\
 & & \quad = \int_{U(1)^n}\md^n\mu_H(\lambda_1,\dots,\lambda_n)\prod_{A'}
  \lambda_{A'}^{-K_{A'}}\cdot\prod_A
  \prod_{p\in\sigma\cap\partial_A\Sigma}\int_{U(1)}\md\mu_H(\lambda_p)
  \delta(\lambda_p,\lambda_A)t_{\sigma,k}(\lambda).\nonumber
\end{eqnarray}
In the last step it is written as an integral of the two functions
multiplied with the delta distribution $\delta({\cal
  C})$. Eq.~(\ref{etab}) leads to the second rigging map
$$
 \eta_2\colon\Psi\to\Psi'\quad,\quad t_{\sigma,k}\mapsto\ket{K_A
   :=\sum_{p\in\sigma\cap\partial_A\Sigma}k_p},
$$
which has to be extended anti-linearly, implementing the boundary
constraints.

The composition of the two maps, $\eta_2\circ\eta_1:\Phi\to\Psi'$,
cannot be used as a rigging map $\eta\colon\Phi\to\Phi'$ because it
has the wrong domain as its image. It can, however, easily be extended
to such a map by interpreting $\sigma$ (the labels of functions in
$\Psi$) as $V(\gamma)\cap\partial\Sigma$ for a graph $\gamma$ labeling
a function
in $\Phi$. This leads us to the rigging map
$$
 \eta\colon T_{\gamma,k}\mapsto\ket{K_A= \sum_{v\in
       V(\gamma)\cap\partial_A\Sigma} k_v}
$$
by extending anti-linearly. Now $\ket{K_1,\ldots,K_n}$ is interpreted
as a distribution in $\Phi'$ analogously to Eq.~(\ref{etab}):
$$
 \ket{K_1,\ldots,K_n}(T_{\gamma,k}):=\prod_A \delta_{K_A,\sum_{v\in
     V(\gamma)\cap\partial_A\Sigma}k_v}.
$$
This rigging map solves both constraints ${\cal F}$ and ${\cal C}$ at
once by incorporating both delta expressions. Moreover, it is real and
positive: $(\eta\phi_1)(\phi_2)=\overline{(\eta\phi_2)(\phi_1)}$ and
$(\eta\phi_1)(\phi_1)\geq 0$ for $\phi_1,\phi_2\in\Phi$. Finally,
$\eta$ commutes with physical observables $O$ by construction
of the observables $\hat{p}^A$ and $\hat{\eta}_A$ via $\kappa_p$ and
$\lambda_p$:
$$
  (\eta\phi_1)(O\phi_2)=(\eta O^*\phi_1)(\phi_2). 
$$
The inner product in the solution space $\eta(\Phi)$ is
given by
$$
  \langle \eta(T_{\gamma,k}),\eta(T_{\gamma',k'})\rangle_{phys}=(\eta
  (T_{\gamma,k})) (T_{\gamma',k'})=\prod_A \delta_{\sum_{v\in
      V(\gamma)\cap\partial_A\Sigma}k_v,\sum_{v'\in
      V(\gamma')\cap\partial_A\Sigma}k'_{v'}}.
$$
The same properties are fulfilled for the partial rigging maps $\eta_1$
and $\eta_2$ yielding the inner products (\ref{innprodmono})
and (\ref{innprod}).

\section{Quantum Symmetry Reduction}

The application of the general quantum symmetry reduction scheme to
the case treated here has already been carried out in Section 4.2 of
Ref.~\cite{SymmRed}, and it suffices to recall the main results and to
compare with the methods of the present paper.

This framework implements a symmetry reduction procedure at the
quantum level, i.e., the theory is spin network quantized first
followed by singling out symmetric states which are distributional and
represented by one-dimensional spin networks in case of spherical
symmetry. The results for spherically symmetric electromagnetism are
the following: There is one degree of freedom given by the electric
charge. The physical Hilbert space is spanned by states $\ket{K}$
exactly as in the preceding section.  Recall that in the interpretation of
$BF$-theory as spherically symmetric electromagnetism there is only
one independent boundary component leading to $n=1$ in the formulae
above. Moreover, the observables are represented in the same way as
derived here yielding the same quantum theory. Furthermore, the
application of a symmetric spin network state, which is a generalized
state in $\Phi'$, on a non-symmetric one
\begin{equation}\label{symm}
 \sigma_g((\eta_B)^K)(T_{\gamma k})=\beta_{\gamma,k}^g\delta_{K,\pi(k)}
\end{equation}
is reminiscent of the rigging map $\eta$.  Here $(\eta_B)^K$ is a
one-dimensional spin network with charge $K$ in the radial manifold
$B$, $g$ is the magnetic charge, $\beta_{\gamma,k}$ a phase factor, and
$\pi(k)$ is a labeling of a one-dimensional spin network projected
from the labeling $k$ which yields for gauge invariant spin networks
the charge (see Ref.~\cite{SymmRed} for details).  Note however that
there appears no rigging map in that paper: In general there will be
no constraint implementing the symmetry reduction as the constraint
${\cal F}$ here.  Therefore, no rigging map, which solves gauge as
opposed to symmetry conditions, is needed. Indeed the methods for
tackling symmetry developed in Ref.~\cite{SymmRed} are quite different
from those to deal with gauge. The present paper shows that in the
electromagnetic example both these methods apply and lead to the same
quantum theory.

In particular, both approaches manage to reduce the infinitely many
degrees of freedom of the non-symmetric field theory to only one in
quite different ways.

There are two main advantages of the general method: First, magnetic
charge is included there from the outset leading to superselection
sectors labeled by the magnetic charge.  In the $BF$-theory approach
we could implement magnetic charge by coupling one in the $BF$-action.
However, this had to be done by hand, whereas all sectors arise
directly in the quantum symmetry reduction.  The constraint reduction
in spin network quantization then had to be performed by evaluating
spin network states in connections of the form
$\omega_a=\omega_a^{(g)}+\partial_a l$, where $\omega_a^{(g)}$ is a
fixed connection with magnetic charge $g$, leading again to the same
boundary spin networks. The magnetic charge then appears only in a
phase factor which is given by the value of a spin network in the
fixed connection $\omega^{(g)}$ and which depends on its geometry
(compare Eq.~(\ref{symm})).

The second, and more important advantage of the quantum symmetry
reduction is that it is a general procedure which applies to any
theory with compact and semisimple (up to $U(1)$-factors) gauge group
and any compact symmetry group (the condition of compactness can be
relaxed). In particular, it applies to symmetry reduction of general
relativity in the real Ashtekar formulation (the gauge group being
$SU(2)$) which was our main motivation to develop that procedure.

The treatment in the present article also allows a comparison of spin
network techniques with the standard Fock space methods used in
Ref.~\cite{Boundary,Vertex} in the context of $BF$-theory. If we do not
impose the surface constraints to provide a more direct comparison
with these two articles, we can use the boundary theory obtained in
Subsection \ref{s:boundary}. The Hilbert space of boundary spin
network functions obtained there has some advantages over the Fock
space quantization: The full group of boundary diffeomorphisms can be
represented, and operators creating charge do not have to be normal
ordered, but they are well defined from the outset in the spin network
context. Thus we see that the spin network representation is well
suited for the kinematical sector of $BF$-theory.

\section*{Acknowledgements}

The author is grateful to H.~Kastrup for discussions and helpful
comments. He also thanks the DFG-Graduierten-Kolleg ``Starke und
elektroschwache Wechselwirkung bei hohen Energien'' for a PhD
fellowship.

\section*{Appendices}

\begin{appendix}
\renewcommand{\theequation}{\thesection.\arabic{equation}}
\setcounter{equation}{0}

\section{Dimensions}
\label{a:units}

In the present article we use electromagnetic dimensions which are
unconventional, but analogous to the geometrical ones used in the
gravitational part of the theory. This means that coordinates as well
as the $U(1)$-connection $\omega_a$ are dimensionless. The electric
field $\epsilon^a$ integrated over a surface yields the enclosed
charge, and therefore it should carry the dimension of electric
charge. (We reserve the letters $A$ and $E$ for the respective
gravitational fields, although they will not appear in this paper. For
the electromagnetic fields we use $\omega$, $\epsilon$, and $\mu$ as
in Ref.~\cite{SphKlEM}.)

A Liouville form with the dimension of an action is given by
$$
 \frac{q}{\alpha}\int_{\Sigma}\md^3x\epsilon^a\delta\omega_a
$$
where $\alpha$ is a dimensionless constant which fixes the norm of the
electromagnetic part of the action, and $q$ is a unit of electric
charge providing the correct dimension of an action. This leads to the
symplectic structure
\begin{equation}\label{Poisson}
 \{\omega_a(x),\epsilon^b(y)\}=\frac{\alpha}{q}\delta^b_a\delta(x,y).
\end{equation}

Up to now we have two constants, $\alpha$ and $q$, which provide the
norm and the dimension, respectively, of the action. We can fix one of
them to obtain a theory with only one undetermined parameter. This
will be done by choosing $q$ in such a way that $\alpha=q^2\hbar^{-1}$
becomes a fine structure constant, and thereby the only parameter.
This leads to the commutator
$[\hat{\omega}_a(x),\hat{\epsilon}^b(y)]=iq\delta^b_a\delta(x,y)$ in
a quantum theory.

\section{Classical Reduction}
\label{a:reduct}

Here we recall the main formulae from Ref.~\cite{SphKlEM} which are used in
the present paper. The basic fields are the electric field
$\epsilon^a$ with density weight $1$ and the $U(1)$-connection
$\omega_a$ which are conjugate to one another. The Gau\ss\ constraint
reads $\partial_a\epsilon^a\approx0$. Symmetry reduction is done by
imposing the restrictions
\begin{equation}
 (\epsilon^x,\epsilon^{\kt},\epsilon^{\kp})=(\epsilon(x,t),0,0)
\end{equation}
and
\begin{equation}
 (\omega_x,\omega_{\kt},\omega_{\kp})=(\omega(x,t),0,0)
\end{equation}
in spherical coordinates $(x,\kt,\kp)$ provided by the given
$SO(3)$-action on the spacelike section $\Sigma$. Here we demand that
there is no magnetic charge. If we use the electric flow
$p:=4\pi\epsilon$, we obtain the two conjugate fields
$(\omega,q\alpha^{-1}p)$ on a radial manifold subject to the Gau\ss\
constraint $p'\approx0$.

For simplicity we restrict $\Sigma$ to be simply connected and to be
either the wormhole manifold $\R\times S^2$, which is the case for a
Reissner-Nordstr\o m black hole, or
$\R^3\backslash\{0\}\cong\R^+\times S^2$, which simulates the presence
of a non-dynamical point charge in the origin. These manifolds are
most interesting in the context of spherical symmetry, but are
generalized slightly in the $BF$-theory approach. Due to simple
connectedness and vanishing of the magnetic field we have
$\omega_a=\partial_a l$ with a function $l\colon\Sigma\to\R$. Symmetry
reduction implies that $l$ is spherically symmetric, i.e., locally
constant on the boundary.

After solving the Gau\ss\ constraint, conjugate variables on the reduced
phase space are found to be $p$, which is constrained to be constant,
and $\Phi:=-\int\md x\omega$. The reduced Hamiltonian accounting for
boundary dynamics is $H_{red}=q\alpha^{-1}pU$ with a prescribed
function $U(t)$ which is the value of the Lagrange multiplier of the
Gau\ss\ constraint at infinity.

\section{{\boldmath$U(1)$}-Spin Networks}
\label{a:U1spin}

To fix our notation we present in this appendix the basic definitions
of $U(1)$-spin networks. Due to Abelianess and the simple
representation theory of $U(1)$ they are more easy to deal with than
$SU(2)$-spin networks. They also appeared in Ref.~\cite{MaxwellSpinnet}.

The irreducible representations of the Abelian $U(1)$ are all
one-dimensional and given by $\rho^k\colon U(1)\to \C^*,g\mapsto g^k$
for all $k\in\Z$. The dual representation of $\rho^k$ is given by
$\rho^{-k}$, and the tensor product of two representations is
$\rho^{k_1}\otimes\rho^{k_2}=\rho^{k_1+k_2}$. A $U(1)$-spin network is
a graph $\gamma$ with a labeling $k\in(\Z\backslash\{0\})^{E(\gamma)}$
of its edge set $E(\gamma)$ with irreducible, non-trivial
$U(1)$-representations. Since the representations are not self-dual,
an inverted edge has to be labeled with the dual representation:
$k_{e^{-1}}=-k_e$. Contrary to the case of $SU(2)$-spin networks, we do
not need contractors in the vertex set $V(\gamma)$, because
intertwiners of $U(1)$-representations are unique up to a constant. If
we do not restrict to gauge invariant spin networks, a coloring of the
vertices which determines the transformation of the spin network under
gauge transformations in that vertex can be computed from the edge
labeling by $k_v=\sum_{e\in E(\gamma)}\sgn(v,e)k_e$ where $v$ is a
vertex of $\gamma$ and $\sgn(v,e)$ is defined to be $1$ if $e$ is an
edge starting at $v$, $-1$ if $e$ ends in $v$, and $0$ otherwise,
i.e., if $v$ is not contained in $e$. Given a graph $\gamma$ with
edge labeling $k$ we can form the spin network state as a function on
the space of generalized $U(1)$-connections \cite{ALMMT} given by
$$
T_{\gamma k}(\omega):=\prod_{e\in E(\gamma)}\rho^{k_e}(\omega(e)),
$$
which transforms under a gauge transformation $g\colon\Sigma\to
U(1)$ by multiplication with the $U(1)$-element $\prod_{v\in
  V(\gamma)}\rho^{k_v}(g(v))$.  Of course, gauge invariant
spin networks are obtained if in all vertices we have $k_v=0$.

The basic operators are multiplication by a holonomy
$\eta_e(\omega):=\exp i\int_e\md t\dot{e}^a\omega_a$ along an edge $e$
which changes the edge labeling by $k_e\mapsto k_e+1$, and a
derivative operator which is the invariant vector field
$i\kappa_e(\eta_e):=i\eta_e\pd{\eta_e}$ on $U(1)$. Again for $U(1)$
being Abelian these operators are much simpler than their analogs in
$SU(2)$.  Spin network states are eigenvectors of $\kappa_e(\eta_e)$
with eigenvalue $k_e$: $\kappa_e(\eta_e)T_{\gamma k}=k_e T_{\gamma
  k}$.

\end{appendix}


\begin{thebibliography}{1}

\bibitem{SymmRed}
M.\ Bojowald and H.~A.\ Kastrup,
\newblock Quantum Symmetry Reduction for Diffeomorphism Invariant Theories of
  Connections,
\newblock hep-th/9907042, to appear in {\em Class.\ Quantum Grav.}

\bibitem{Horo:BF}
G.~T.\ Horowitz,
\newblock Exactly Soluble Diffeomorphism Invariant Theories,
\newblock {\em Commun.\ Math.\ Phys.} 125 (1989) 417--437

\bibitem{SphKlEM}
T.\ Thiemann,
\newblock The reduced phase space of spherically symmetric Einstein-Maxwell
  theory including a cosmological constant,
\newblock {\em Nucl.~Phys.~B} 436 (1995) 681--720

\bibitem{BlauThom:BF}
M.\ Blau and G.\ Thompson,
\newblock Topological Gauge Theories of Antisymmetric Tensor Fields,
\newblock {\em Ann.\ Phys.} 205 (1991) 130--172

\bibitem{Wu:BFRand}
S.\ Wu,
\newblock Topological Quantum Field Theories on Manifolds with a Boundary,
\newblock {\em Commun.\ Math.\ Phys.} 136 (1991) 157--168

\bibitem{Boundary}
A.~P.\ Balachandran and P.\ Teotonio-Sobrinho,
\newblock The Edge States of the $BF$ System and the London Equations,
\newblock {\em Int.\ J.\ Mod.\ Phys.\ A} 8 (1993) 723--752

\bibitem{Vertex}
A.~P.\ Balachandran and P.\ Teotonio-Sobrinho,
\newblock Vertex Operators for the $BF$ System and its Spin-Statistics
  Theorems,
\newblock {\em Int.\ J.\ Mod.\ Phys.\ A} 9 (1994) 1569--1629

\bibitem{BFSU}
V.\ Husain and S.\ Major,
\newblock Gravity and $BF$ theory defined in bounded regions,
\newblock {\em Nucl.\ Phys.} B 500 (1997) 381--401, [gr-qc/9703043]

\bibitem{ALMMT}
A.\ Ashtekar, J.\ Lewandowski, D.\ Marolf, J.\ Mour\~ao, and T.\ Thiemann,
\newblock Quantization of diffeomorphism invariant theories of connections with
  local degrees of freedom,
\newblock {\em J.\ Math.\ Phys.} 36 (1995) 6456--6493, [gr-qc/9504018]

\bibitem{MaxwellSpinnet}
A.\ Corichi and K.\ Krasnov,
\newblock Loop Quantization of Maxwell Theory and Electric Charge Quantization,
\newblock {\em Mod.\ Phys.\ Lett.} A13 (1998) 1339--1346, [hep-th/9703177]

\end{thebibliography}
\end{document}